\documentclass[aps,prl,epsfig,showpacs,floats,twocolumn,amssymb,amsmath,floatfix]{revtex4-1}

\usepackage{amsmath}
\usepackage{graphicx}
\usepackage{xspace}

\begin{document}

\title{Enhanced Diffusion of Enzymes that Catalyze Exothermic Reactions}

\author{Ramin Golestanian}
\email[]{ramin.golestanian@physics.ox.ac.uk}
\affiliation{Rudolf Peierls Centre for Theoretical Physics, University of Oxford, Oxford OX1 3NP, UK}

\date{\today}

\begin{abstract}
Enzymes have been recently found to exhibit enhanced diffusion due to their catalytic activities. A recent experiment [C. Riedel {\it et al.}, Nature {\bf 517}, 227 (2015)] has found evidence that suggests this phenomenon might be controlled by the degree of exothermicity of the catalytic reaction involved. Four mechanisms that can lead to this effect, namely, self-thermophoresis, boost in kinetic energy, stochastic swimming, and collective heating, are critically discussed, and it is shown that only the last two could be strong enough to account for the observations. The resulting quantitative description is used to examine the biological significance of the effect.
\end{abstract}

\maketitle

\emph{Introduction.}---A most fascinating aspect of the nonequilibrium processes in living cells is active transport \cite{Marchetti:2013}. The basic units of these processes, which could be in the form of carrying cargo or sliding actin fibres against one another, are motor proteins that convert chemical energy directly into useful mechanical work, amidst dominant thermal fluctuations at the nano-scale \cite{Svoboda:1993}. Recent \emph{in vitro} studies of mixtures of motors and filaments have revealed their remarkable ability to self-organize into dynamic meso-scale structures that resemble those observed in living cells \cite{Bausch:2010,Sanchez:2012,Nagai:2012}. Much less is known about the nature of the nonequilibrium activity of non-cytoskeletal elements and how they self-organize in living cells.

It has been recently reported that enzymes undergo {\em enhanced diffusion}, i.e. diffusive motion with an effective diffusion coefficient $D_{\rm eff}$ that is larger than its equilibrium value $D_0$, as a result of their catalytic activity \cite{Sen-1,Sen-2}. Considering the enzyme as a sphere of radius $R$ in a medium with viscosity $\eta$ and temperature $T$, the Stokes-Einstein relation $D_0=k_{\rm B} T/\zeta$ gives us the equilibrium diffusion coefficient of the enzyme, where $\zeta=6 \pi \eta R$ is its friction coefficient. The additional, nonequilibrium, contribution to the diffusion coefficient, $\Delta D=D_{\rm eff}-D_0$, is found to be proportional to the net rate (or speed) of the catalytic reaction. The rate has the characteristic Michaelis-Menten form $k=k_e S/(K_M+S)$, where $S$ is the substrate (i.e. reactant) concentration, $K_M$ is the Michaelis constant, and $k_e$ is the enzyme reaction rate. Remarkably, $\Delta D$ has the same order of magnitude as the equilibrium diffusion coefficient; typically a fraction of it. It has also been observed that there is a strong correlation between the degree of exothermicity of the catalytic reaction and the enhancement in the effective diffusion coefficient of enzymes \cite{Bustamante-1}. In this Letter, I discuss and critically examine various mechanisms that can lead to enhanced diffusion for catalytically active enzymes.

\emph{Self-phoresis.}---Any colloidal particle that actively generates nonequilibrium phoretic flow in its vicinity exhibits enhanced effective diffusion at time scales longer than its orientational persistence time $1/D_r$, where $D_r$ is the rotational diffusion coefficient \cite{Golestanian:2009b}. Enzyme activity has the right ingredients to lead to enhanced diffusion via self-diffusiophoresis, which takes advantage of gradients in the concentrations of the chemicals involved in the reaction \cite{Golestanian:2005}. However, this contribution is not sensitive to the degree of exothermicity of the reaction. The appropriate mechanism that could account for such an effect is self-thermophoresis \cite{Golestanian:2007,Jiang:2010}. The heat released from the chemical reaction during each catalytic cycle, $Q$, leads to a temperature difference $\Delta T \simeq k Q/(\kappa R)$ across the enzyme, where $\kappa$ is the thermal conductivity of the medium. For catalase $k=5 \times 10^4\,$s$^{-1}$, $Q=40 \,k_{\rm B} T$, and $R=4\,$nm, which gives $D_0=55\,\mu{\rm m}^2{\rm s}^{-1}$ and $D_r=2.6 \times 10^6 \, {\rm s}^{-1}$. Using these values and $\kappa \sim 0.6 \,{\rm W/(m \cdot K)}$ for water, we obtain $\Delta T \sim 10^{-6}$K. The self-propulsion velocity is estimated as $V_{\rm st} \sim D_0 S_T \Delta T/R \sim D_0 S_T k Q/(\kappa R^2)$ \cite{Golestanian:2007}, where $S_T$ is the Soret coefficient of the enzyme. Using $S_T \simeq 0.02\,{\rm K}^{-1}$ \cite{putnam}, we find $V_{\rm st} \sim 10^{-4}\,\mu{\rm m} \, {\rm s}^{-1}$. The correction to effective diffusion coefficient due to self-thermophoresis is $\Delta D \simeq V_{\rm st}^2/D_r$, for which we find $\Delta D \simeq 10^{-14}\,\mu{\rm m}^2 {\rm s}^{-1}$, and consequently $\Delta D/D_0 \simeq 10^{-16}$. This is fifteen orders of magnitude too small to account for the observations.

\emph{Boost in Kinetic Energy.}---The authors of Ref. \cite{Bustamante-1} propose a scenario in which the heat released from the chemical reaction during each catalytic cycle is channeled into a boost in the translational velocity of the enzyme. This is not envisaged to be mediated through an effective temperature increase following the release of heat. Here I reproduce their analysis using a slightly different derivation, to highlight the essence of the proposed mechanism. Consider the enzyme to be a particle of mass $m$ whose stochastic motion satisfies Newton's equation $m \frac{d^2 {\bf r}}{d t}+\zeta \frac{d {\bf r}}{d t}={\bf f}(t)$, where ${\bf f}(t)$ is a random force. The relative significance of the inertial and the dissipative terms in the equation of motion is characterized by the time scale $\tau=m/\zeta$. Using $m=\frac{4 \pi}{3} R^3 \rho_p$ with a typical protein mass density $\rho_p=1.4 \times 10^3\, {\rm kg}/{\rm m}^3$ and  $\eta=1 \times 10^{-3}\,$Pa$\cdot$s for viscosity of water at room temperature, we find $\tau=5\,$ps. Invoking an elegant mathematical trick that Langevin used in his original 1908 paper \cite{Langevin}, we can write the equation of motion as
\begin{equation}
\left[\tau \frac{d}{d t}+1 \right] D_{\rm eff}(t)=\frac{2}{3 \zeta} \; {\cal E}(t),\label{eq:Langevin}
\end{equation}
where $D_{\rm eff}(t)=\frac{1}{6} \frac{d}{d t} \langle {\bf r}(t)^2\rangle$ is by definition the effective diffusion coefficient and ${\cal E}(t)=\langle \frac{m}{2} \left(\frac{d {\bf r}}{d t}\right)^2\rangle$ is the average kinetic energy of the enzyme. Here we have assumed a separation of time scales between the random thermal kicks that the enzyme receives from the medium and the catalytic cycle, and the averaging is performed over the thermal kicks. We can write ${\cal E}(t)=\frac{3}{2} k_{\rm B} T+\gamma Q h(t)$ where $\gamma$ represents the fraction of the released thermal energy that is converted into the translational boost and $h(t)$ is a series of spikes of width $\tau_b$ that appear stochastically at a rate $k$, through a Poisson process. Here, $\tau_b$ is the relaxation time of the boost, which depends on the specific process that generates it. It is reasonable to assume that $\tau_b \approx \tau$. The boost mechanism proposed in Ref. \cite{Bustamante-1} involves asymmetric excitation of compressional waves along the enzyme that propagate to the interface with water and trigger a pressure wave that leads to a back-reaction on the enzyme itself, giving it a mechanical boost. No evidence is provided in Ref. \cite{Bustamante-1} as to why the energy is not randomly partitioned between a large number of possible channels (owing to the large number of degrees of freedom or normal modes), which would result in $\gamma \ll 1$, and subsequently dissipated, as opposed to being channeled to a small number of modes (corresponding to $\gamma \sim 1$). Time averaging gives $\overline{{\cal E}}=\frac{3}{2} k_{\rm B} T+\gamma Q k \tau_b$, and consequently
\begin{equation}
\left.\frac{\Delta D}{D_0}\right|_{{\rm Ref. [8]}}=\frac{2}{3}\, \frac{\gamma Q}{k_{\rm B} T} \,k \tau_b,   \label{eq:Presse}
\end{equation}
through Eq. (\ref{eq:Langevin}). Using the above estimate for $\tau_b$ and the values for $k$ and $Q$ corresponding to catalase, we obtain $k \tau_b=2.5 \times 10^{-7}$ and $\Delta D/D_0=\gamma \times 10^{-5}$ from Eq. (\ref{eq:Presse}). Even with the (unrealistic) maximum value of $\gamma=1$, our estimate from Eq. (\ref{eq:Presse}) is four orders of magnitude too small to account for the observations.

\emph{Stochastic Swimming.}---We could examine various hydrodynamic effects that might contribute towards such a behaviour. Substrate binding could change the shape of the enzyme, and consequently its friction coefficient. Since this process does not require energy input, however, it cannot be the cause of a nonequilibrium phenomenon. Such conformational changes are often relatively small, and will more likely lead to an increase in size that would result in a decrease in diffusion coefficient, rather than the other way around. Moreover, it is not clear why such an effect could lead to universal trends---as it will depend on specific cases---and how it can correlate with exothermicity.

As an alternative scenario, it is possible that the catalytic cycle induces conformational changes in the enzyme that lead to stochastic swimming \cite{Golestanian:2008b}. The amplitude of these deformations is typically much smaller than the size of the enzyme, e.g. when they arise from mechanochemical coupling of electrostatic nature \cite{Golestanian:2010} (analogous to phosphorylation) or structural changes due to ligand binding \cite{Sakaue:2010}. However, local heat release could have the possibility to transiently disturb the relatively more fragile tertiary structure of the folded protein \cite{Wolynes:2003} or the state of oligomerization of the enzyme \cite{oligo}, and produce an amplitude $b$ that is a fraction of the size $R$.

To calculate the contribution of such conformational changes to effective diffusion coefficient, we use a simple model in which the conformational change is described by one degree of freedom $L(t)$ representing elongation of the structure along an axis defined by a unit vector $\hat{\bf n}(t)$. To achieve directed swimming, we need at least two degrees of freedom to incorporate the coherence needed for breaking the time-reversal symmetry at a stochastic level, and we know that realistic conformational changes must involve many degrees of freedom. The randomization of the orientation, described via $\left\langle \hat{\bf n}(t) \cdot \hat{\bf n}(t')\right\rangle=e^{-2 D_r |t-t'|}$, will turn the directed motion into enhanced diffusion over the time scales longer than $1/D_r$. Since the same can be achieved through reciprocal conformational changes described by one compact degree of freedom, I will adopt this simpler form. The stochastic motion of the enzyme can be described by the Langevin equation ${\bf v}(t) \simeq \alpha \left(\frac{d}{d t} L\right) \, \hat{\bf n}(t)+ \boldsymbol{\xi}(t)$ where $\alpha$ is a numerical pre-factor that depends on the geometry of the enzyme \cite{note-2,note-3} and $\boldsymbol{\xi}(t)$ is the Gaussian white noise that will give us the intrinsic translational diffusion coefficient $D_0$.

\begin{figure}[b]
%\vskip-4mm
\begin{center}
\includegraphics[width=0.75\columnwidth]{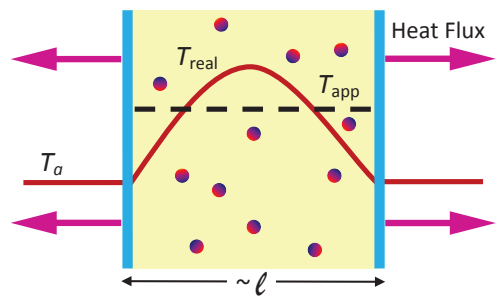}
\end{center}
\vskip-4mm
\caption{(color online.) Schematic illustration of the essence of Newton's law of cooling. The heat generated in the bulk of a chamber (that is in contact with the environment that has a fixed ambient temperature $T_a$), and lost in the form of heat flux through the boundaries leads to the temperature profile $T_{\rm real}$ (solid line), which is approximated by the average value over the distance $\sim \ell$, $T_{\rm app}$ (dashed line). This approximation works best when there is a separation of length scales.}
   \label{fig:schem}
\end{figure}

\begin{figure*}[t]
\begin{center}
\includegraphics[width=2.0\columnwidth]{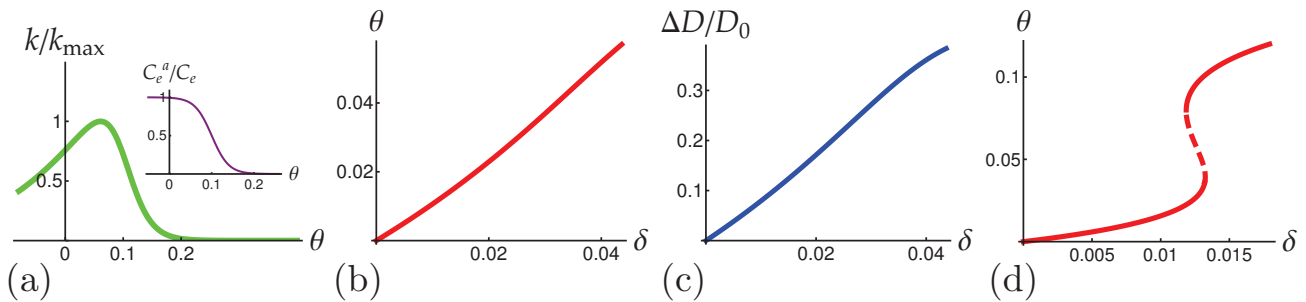}
\end{center}
\caption{(color online.) (a) The effective rate of catalytic reaction as a function of the reduced temperature $\theta$ for $\theta_d=0.1$, $g=50$, and $\epsilon=7$. The inset shows the fraction of catalytically active enzymes as a function of $\theta$ in each case. (b) The effective temperature of the catalytically active medium as a function of the coupling strength $\delta$. (c) The corresponding relative increase in diffusion coefficient. (d) The effective temperature for $\epsilon=30$, with the dashed region being unstable.}
\label{fig:plots}
\end{figure*}

We describe the combined mechanochemical cycle using a two-step process, which takes the enzyme from its free state to the reaction stage that is followed by the deformation with rate $k$, and a relaxation back to its native state with rate $k_r$. This is a simplification of a more realistic model with three states (free, substrate-bound, and reacted-deformed) and $k$ is to be understood as the combined catalytic rate that has the Michaelis-Menten form as defined above. In stationary state, a master equation formulation can be used to calculate the elongation speed autocorrelation function as $\left\langle \frac{d}{d t} L(t) \cdot \frac{d}{d t'} L(t')\right\rangle=2 b^2 \left(\frac{k k_r}{k+k_r}\right) [\delta(t-t')-\frac{1}{2} (k+k_r)e^{-(k+k_r) |t-t'|}]$. By combining this with the orientation auto-correlation, we can calculate the effective diffusion coefficient of the enzyme, which gives the following correction
\begin{equation}
\Delta D=\frac{1}{3} \alpha^2 b^2 \left(\frac{k k_r}{k+k_r}\right) \frac{2 D_r}{2 D_r+k+k_r}.\label{eq:DeltaD-swimming-1}
\end{equation}
Even for the fastest enzymes, we typically have $k_r \approx D_r \gg k$. Using an upper bound of $b \lesssim R$, we can approximate Eq. (\ref{eq:DeltaD-swimming-1}) as $\Delta D \approx k R^2$. For catalase, we obtain $\Delta D \approx 1\,\mu{\rm m}^2 {\rm s}^{-1}$, which gives an upper bound of $\Delta D/D_0 \approx 10^{-2}$. This is one order of magnitude smaller than the observed values.

\emph{Collective Heating.}---In Ref. \cite{Bustamante-1}, a calculation similar to what we have above to estimate the relative change in temperature across the enzyme, $\Delta T$, is used to argue that heating of the environment by the enzyme is negligible. This estimate, however, is only correct for an isolated enzyme. In practice, an experiment is performed on a solution with a finite concentration of enzyme, $C_e$. For such a sample, the substrate is consumed at the rate (per unit volume) of $k_e S C^a_e/(K_M+S)$, where $C^a_e$ is the concentration of the catalytically active enzymes. The exothermic catalytic reaction generates thermal energy $Q$ per turnover cycle at the location of each active enzyme, which then diffuses through the sample container and escapes via the boundaries. Due to the large number of heat producing enzymes (of the order of Avogadro's number) there will be a significant build-up of thermal energy in the sample container. To see this, let us define a length scale $\ell$ that describes the characteristic distance heat needs to diffuse until it can exit. $\ell$ is typically set by the smallest length scale in the geometry of the sample container. We can estimate a characteristic heat diffusion time $\tau_{h}(\ell)=\ell^2/\chi$, using the thermophoretic conductivity $\chi$. For water at room temperature, we have $\chi \simeq 10^5 \,\mu{\rm m}^2 {\rm s}^{-1}$. For $\ell=10\,$mm, it takes $\tau_h=1000\,$s for the heat released from each enzyme during each catalytic cycle to leave the container. Given $C_e=1\,$nM \cite{Bustamante-1}, during this time $10^{20}$ units of $Q$ will have been released into the chamber (assumed to have a volume $\sim \ell^3$); this is $10\,$J of thermal energy.

The heat diffusion equation is written as
\begin{equation}
\chi^{-1} \partial_t T- \nabla^2 T=\frac{Q}{\kappa} \cdot \frac{k_e S C^a_e}{K_M+S}-\frac{1}{\ell^2}(T-T_a).\label{eq:T1}
\end{equation}
The right hand side of Eq. (\ref{eq:T1}) contains a source term that couples the catalytic reaction to the production of heat, and a sink term that approximates the heat loss through the boundaries by a bulk term in the form of Newton's law of cooling \cite{Newton:1701}, where $T_a$ is the ambient temperature.
This term is written in terms of the length scale $\ell$ that is described above. This approximation has been widely used in the combustion literature to allow the temperature-dependent nonlinearities in the source term to be captured in a manner that does not depend on the geometric specificities of each experiment \cite{Zeldovich}. Figure \ref{fig:schem} summarizes the essence of this approximation.

Two quantities in the source term of Eq. (\ref{eq:T1}) have significant dependence on temperature, which we will represent using $\theta=(T-T_a)/T_a$. The first quantity is the turnover rate, which we can assume to have an Arrhenius form of $k_e=k^*_0 e^{-E_a/k_{\rm B}T}$, where $E_a$ is the activation energy. The rate can be rewritten as $k_e=k_0 e^{\epsilon \theta /(1+\theta)}$, where $\epsilon=E_a/k_{\rm B}T_a$ and $k_0=k^*_0 e^{-\epsilon}$. The second quantity is the concentration of active enzymes. Since enzymes are proteins, increase in temperature will eventually denature them upon approaching the denaturation temperature, which we denote by $\theta_d$. We can use a simple two-state model to account for the denaturation, which yields $C^a_e=C_e/\left[e^{g (\theta-\theta_d)}+1\right]$, where the parameter $g$ controls the sharpness of the transition. The two effects enter the heat source term in Eq. (\ref{eq:T1}) via the product $k_e C^a_e$, whose temperature dependence is shown in Fig. \ref{fig:plots}(a), for $g=50$, $\theta_d=0.1$ (that corresponds to $T_d=330$K for $T_a=300$K), and $\epsilon=7$, which is a typical value for the activation energy for enzymes, such as catalase \cite{Sizer:1944}. The plot shows the two standard regimes of initial increase in the effective rate due to the Arrhenius temperature dependence and the sudden decline due to denaturation. The peak will sharpen with increasing activation energy $\epsilon$. The inset of Fig. \ref{fig:plots}(a) shows the fraction of active enzymes as a function of temperature.

Writing the temperature dependencies in Eq. (\ref{eq:T1}) explicitly, we find
\begin{equation}
\tau_h \partial_t \theta-\ell^2 \nabla^2 \theta=\delta \; \frac{e^{\epsilon \theta /(1+\theta)}}{e^{g (\theta-\theta_d)}+1}-\theta,\label{eq:T2}
\end{equation}
where
\begin{equation}
\delta=\frac{Q \ell^2}{\kappa T_a} \cdot \frac{k_0 S C_e}{K_M+S},\label{eq:delta-def}
\end{equation}
emerges as a single dimensionless parameter that controls the strength of collective heating. Assuming a uniform profile, we can find the stationary state temperature of the system as a function of $\delta$ by setting the right hand side of Eq. (\ref{eq:T2}) to zero. The result is shown in Fig. \ref{fig:plots}(b). Note that the temperature dependence of the kinetic rate provides such a sensitive positive feedback mechanism that could lead to unrealistically high temperatures, easily above enzyme denaturation and even above boiling temperature of water (for a case such as catalase with the large value of $Q=40 \,k_{\rm B} T$). The presence of denaturation provides a negative feedback mechanism that ensures such dramatic temperature increases are cut off.

The temperature increase can be used to calculate the relative increase in diffusion coefficient $\Delta D/D_0$, by taking into account the corresponding variations in the friction coefficient. Denaturation will change the hydrodynamic radius of the protein from its globular (folded) form, with size $R \sim a N^{1/3}$, to its coiled (unfolded) form, with size $R_g \sim a N^{3/5}$, where $a$ is the Kuhn length and $N$ is the polymerization index. This yields $R_g \sim R N^{4/15}$. For catalase, we have $a \simeq 1\,$nm, $N \simeq 100$, $R_g \simeq 14\,$nm. To calculate the diffusion coefficient within our simple two-state model, we need to use the ensemble average of the inverse size, namely, use $1/R$ with the probability $p=1/\left[e^{g (\theta-\theta_d)}+1\right]$ and $1/R_g$ with the probability $1-p$. We can use the following expression for the temperature dependence of the viscosity of water $\eta(\theta)=\eta_0 \exp\left\{\frac{(B/T_a)}{1+\theta-(T_0/T_a)}\right\}$ where $B=579$K, $T_0=138$K, and $\eta_0=2.41 \times 10^{-5}\,$Pa$\cdot$s. The result is plotted in Fig. \ref{fig:plots}(c), showing a similar pattern of behaviour as seen in $\theta$. The values obtained for $\Delta D/D_0$ are of the same order of magnitude as the experimentally observed values, while the actual temperature increase in the solution is relatively modest. Moreover, the trend in Fig. \ref{fig:plots}(c) resembles the experimental results reported in Refs. \cite{Sen-1,Sen-2,Bustamante-1}.

To estimate typical experimental values for $\delta$, we need a typical length scale from the sample container and the smallest conductivity involved in the geometry of the system. For Fluorescence Correlation Spectroscopy (FCS) experiments, a droplet of the solution is placed over a glass slide. Using $\kappa \sim 0.02 \,{\rm W/(m \cdot K)}$ for air, $\ell=5\,$mm, and $C_e=1\,$nM (as used in Ref. \cite{Bustamante-1}), $k_0=5 \times 10^4\,$s$^{-1}$, and $Q=40 \,k_{\rm B} T$, we find $\delta \simeq 0.02$ at substrate saturation. This estimate appears to provide a rough order of magnitude agreement with the observations of Ref. \cite{Bustamante-1}. Note that due to our approximate treatment of the boundary conditions, order unity differences are to be expected when comparing to the experimental results. The experiments in Refs \cite{Sen-1,Sen-2} use higher concentrations of enzymes and similar sample sizes, and thus comfortably fall in the regime described by the collective heating scenario.

\emph{Discussion.}---The collective heating model leads to very specific predictions that could be experimentally tested. The near linear dependence of the relative enhancement in diffusion coefficient on $\delta$ [Fig. \ref{fig:plots}(c)] and the definition of $\delta$ [Eq. (\ref{eq:delta-def})] suggest a linear dependence on the enzyme concentration and a quadratic dependence on the size of the container. In practice, protein denaturation is irreversible, which suggests that recording experimental data over a long time scale would presumably lead to a systematic reduction in the magnitude of the enhanced diffusion, provided the experiment is done under the condition that the substrate concentration is maintained at a constant level.

The heating mechanism can also lead to interesting nonlinear phenomena. While at small values of $\epsilon$, we observe a near-linear dependence of temperature on $\delta$ [Fig. \ref{fig:plots}(b)], upon increasing $\epsilon$ the curve takes an S-shape that develops an instability at sufficiently large values of $\epsilon$, as shown in Fig. \ref{fig:plots}(d). The instability will lead to the formation of waves, which will dissipate in a sealed sample container when all the fuel molecules are consumed, following closely the phenomenology of {\it flames} in combustion \cite{Mercer:1996}. Moreover, collective heating could have synergistic influence on the other mechanisms: while the increase in temperature could facilitate the emergence of large conformational changes in the tertiary structure or the oligomerization state during the enzymatic turnover, phoretic collective heating can lead to further instabilities \cite{rg2012} that could accentuate the degree of fluctuations in the system.

Finally, let us examine whether the total heat generated in a cell could be sufficient to trigger this effect. Considering a cell of size $\ell=10\,\mu$m that is fully packed with enzymes similar to catalase (that gives us $C_e\sim 1\,$mM), we find $\delta \sim 0.01$. While this is an upper limit, it certainly points to a strong possibility that the enhanced diffusion via collective heating could be a contributing factor to non-directed intracellular transport in living cells.

In conclusion, enhanced diffusion of enzymes that catalyze exothermic reactions could be explained by a combination of global temperature increase in the sample container, and possibly enhanced conformational changes that can lead to a hydrodynamic enhancement of effective diffusion coefficient. Self-thermophoresis and boost in kinetic energy as suggested by Ref. \cite{Bustamante-1} are too weak to account for the experimentally measured values of effective diffusion. Although the primary focus of this work has been on enhanced diffusion of enzymes, the theoretical description should be relevant to the study of any class of thermally activated microswimmers.

%Biological relevance of this phenomenon is as yet unclear. However, the effect can be strong enough to affect the intracellular stochastic transport of biomolecules.

\acknowledgements

I have benefitted from discussions with Carlos Bustamante, Krishna Kanti Dey, and Ayusman Sen.


\begin{thebibliography}{10}
\bibitem{note-2}
This form is strictly speaking only correct for shapes with axial symmetry about the axis of deformation. For more realistic shapes, $\alpha$ will be a tensor. Introducing this level of description is an unnecessary complication at this level.
\bibitem{note-3}
For a simple model of two spheres of radii $a_1$ and $a_2$, we have $\alpha=(a_1-a_2)/(a_1+a_2)$. In our order of magnitude estimates, we use an upper limit value of $|\alpha|=1$.
\end{thebibliography}
\end{document}